\documentclass[11pt,preprint]{aastex} 
\usepackage{spr-astr-addons}
\usepackage{url}

\RequirePackage{color}
\begin{document}
\title{The 155-day periodicity of the sunspot area 
fluctuations in the solar cycle 16 is an alias}
\shorttitle{The 155-day periodicity}
\shortauthors{Getko}
\author{R. Getko}
\affil{Astronomical Institute, Wroc{\l}aw University, Kopernika 11, 
51--622 Wroc{\l}aw, Poland\\}
\email{getko@astro.uni.wroc.pl}

\begin{abstract}
The short-term periodicities of the daily sunspot area fluctuations from 
August 1923 to October 1933 are discussed. For these data the correlative 
analysis indicates negative correlation for the periodicity of about $155$ days, 
but the power spectrum analysis indicates a statistically significant peak 
in this time interval. A new method of the diagnosis of an echo-effect in 
spectrum is proposed and it is stated that the 155-day periodicity is 
a harmonic of the periodicities from the interval of $[400,500]$ days.

The autocorrelation functions for the daily sunspot area
fluctuations and for the fluctuations of the one rotation time interval 
in the northern hemisphere, separately for the whole solar cycle 16 
and for the maximum activity period of this cycle do not show differences, 
especially in the interval of $[57, 173]$ days. It proves against the thesis 
of the existence of strong positive fluctuations of the about $155$-day interval 
in the maximum activity period of the solar cycle 16 in the northern hemisphere. 
However, a similar analysis for data from the southern hemisphere indicates that 
there is the periodicity of about $155$ days in sunspot area data in the maximum 
activity period of the cycle 16 only.

\end{abstract}

\section{Introduction}

For about 20 years the problem of properties of short-term changes of solar 
activity has been considered extensively. Many investigators studied the short-term 
periodicities of the various indices of solar activity. 
Several periodicities were detected, but the periodicities about 
155 days and from the interval of $[470, 620]$ days ($[1.3, 1.7]$ years) 
are mentioned most often.
First of them was discovered by \citet{rieg} 
in the occurence rate of gamma-ray flares detected by the gamma-ray
spectrometer aboard the \it{Solar Maximum Mission (SMM)}.
\rm This periodicity was confirmed for other solar flares data and 
for the same time period  \citep{bog, bai7, kile}. It was also 
found in proton flares during solar cycles 
19 and 20 \citep{baic}, but it was not found in the solar flares data
during solar cycles 22 \citep{kile, bai2, ozg}. 

Several autors confirmed above results for the daily sunspot area 
data. \citet{lea} studied the sunspot data from 
1874--1984. She found the 155-day periodicity in data records 
from 31 years. This periodicity is always 
characteristic for one of the solar hemispheres (the southern 
hemisphere for cycles 12--15 and the northern hemisphere for
cycles 16--21). Moreover, it is only present during epochs of
maximum activity (in episodes of 1--3 years).

Similar\quad investigations\quad were\quad carried\quad out\quad by \\
\citet{car2}. They applied the same power spectrum method as 
Lean, but the daily sunspot area data (cycles 12--21) were divided
into 10 shorter time series. The periodicities were searched for 
the frequency interval 57--115 nHz (100--200 days) and for each 
of 10 time series. The authors showed that the periodicity between 
150--160 days is statistically significant during all cycles
from 16 to 21. The considered peaks were remained unaltered 
after removing the 11-year cycle and applying the power spectrum 
analysis. 

\citet{ol8} used the wavelet technique
for the daily sunspot areas between 1874 and 1993. They determined the 
epochs of appearance of this periodicity and concluded that it 
presents around the maximum activity period in cycles 16 to 21.
Moreover, the power of this periodicity started growing at 
cycle 19, decreased in cycles 20 and 21 and disappered after 
cycle 21.

Similar\quad analyses\quad were\quad presented\quad by \\ \citet{pra},
but for sunspot number, solar wind plasma, interplanetary magnetic field 
and geomagnetic activity index $A_p$. During 1964-2000  the sunspot 
number wavelet power of periods less than one year shows a cyclic evolution 
with the phase of the solar cycle.The 154-day period is prominent and its strenth 
is stronger around the 1982-1984 interval in almost all solar wind parameters. 
The existence of the 156-day periodicity in sunspot data were confirmed by 
\citet{kri}. They considered the possible relation between 
the 475-day (1.3-year) and 156-day periodicities. The 475-day (1.3-year) periodicity 
was also detected in variations of the interplanetary magnetic field, 
geomagnetic activity helioseismic data and in the solar wind speed 
\citep{pau, sza, loo, rich}. \citet{pra} concluded that 
the region of larger wavelet power shifts from 475-day (1.3-year) period to 
620-day (1.7-year) period and then back to 475-day (1.3-year). 
The periodicities from the interval $[475, 620]$ days ($[1.3, 1.7]$ years)
have been considered from 1968. \citet{yac}
mentioned a 16.3-month (490-day) periodicity in the sunspot 
numbers and in the geomagnetic data. \citet{bai7} analysed the occurrence rate of major flares 
during solar cycles 19. They found a 18-month (540-day) periodicity 
in flare rate of the norhern hemisphere.
\citet{ichi} confirmed this result for the $H_{\alpha}$
flare data for solar cycles 20 and 21 and found a peak in the power 
spectra near 510--540 days. \citet{aki}
found a 17-month (510-day) periodicity of sunspot groups and their 
areas from 1969 to 1986. These authors concluded that the length of 
this period is variable and the reason of this periodicity is still
not understood. 

\citet{car2} and \\ \citet{ol8} 
obtained statistically significant peaks of power at around 158 days 
for daily sunspot data from 1923-1933 (cycle 16). In this paper the problem 
of the existence of this periodicity for sunspot data from cycle 16
is considered. The daily sunspot areas, the mean sunspot areas 
per Carrington rotation, the monthly sunspot numbers and their fluctuations, 
which are obtained after removing the 11-year cycle are analysed. In Section 2 
the properties of the power spectrum methods are described. In Section 3 a new 
approach to the problem of aliases in the power spectrum analysis is presented. 
In Section 4 numerical results of the new method of the diagnosis of an echo-effect 
for sunspot area data are discussed. In Section 5 the problem of the existence of 
the periodicity of about 155 days during the maximum activity period for sunspot data 
from the whole solar disk and from each solar hemisphere separately is considered.

\section {Methods of periodicity analysis}

To find periodicities in a given time series the power spectrum 
analysis is applied. In this paper two methods are used: 
The Fast Fourier Transformation algorithm with the Hamming window function (FFT) 
and the Blackman-Tukey (BT) power spectrum method \citep{black}.

The BT method is used for the diagnosis of the reasons 
of the existence of peaks, which are obtained by the FFT method. The BT 
method consists in the smoothing of a cosine transform of an 
autocorrelation function using a 3-point weighting average. Such an 
estimator is consistent and unbiased. Moreover, the peaks are 
uncorrelated and their sum is a variance of a considered time series. 
The main disadvantage of this method is a weak resolution of the 
periodogram points, particularly for low 
frequences. For example, if the autocorrelation function is evaluated
for $i=1,\ldots ,1000$, then the distribution points in the time domain 
are: $1000, 500, 400, 333, 250, 167, 154,\ldots$ Thus, it is obvious
that this method should not be used for detecting low frequency 
periodicities with a fairly good resolution. However, because of an 
application of the autocorrelation function, the BT method can be used 
to verify a 'reality' of peaks which are computed using a method giving 
the better resolution (for example the FFT method).

It is valuable to remember that the power spectrum methods should be 
applied very carefully. The difficulties in the interpretation of significant 
peaks could be caused by at least four effects: a sampling of a continuos 
function, an echo-effect, a contribution of long-term periodicities 
and a random noise.

First effect exists because periodicities, which are shorter than the 
sampling interval, may mix with longer periodicities. In result, this 
effect can be reduced by an decrease of the sampling interval between 
observations.

The echo-effect occurs when there is a latent harmonic of frequency 
$\alpha$ in the time series, giving a spectral peak at $\alpha$, and
also periodic terms of frequency $\frac{1}{2}\alpha, \frac{1}{3}\alpha,$
etc. This may be detected by the autocorrelation function
for time series with a large variance.

Time series often contain long-term periodicities, that influence 
short-term peaks. They could rise periodogram's peaks at lower 
frequencies. However, it is also easy to notice the influence of the 
long-term periodicities on short-term peaks in the graphs of the 
autocorrelation functions. This effect is observed for the time 
series of solar activity indexes which are limited by the 11-year cycle.

To find statistically significant periodicities it is reasonable to 
use the autocorrelation function and the power spectrum method with a 
high resolution. In the case of a stationary time series they give similar 
results. Moreover, for a stationary time series with the mean zero the Fourier 
transform is equivalent to the cosine transform of an autocorrelation function
\citep{and}. Thus, after a comparison of a  periodogram with an 
appropriate autocorrelation function one can detect peaks which 
are in the graph of the first function and do not exist in the graph
of the second function. The reasons of their existence could be 
explained by the long-term periodicities and the echo-effect. Below method 
enables one to detect these effects.
 \begin{figure}
 \centerline{\includegraphics[width=20pc]{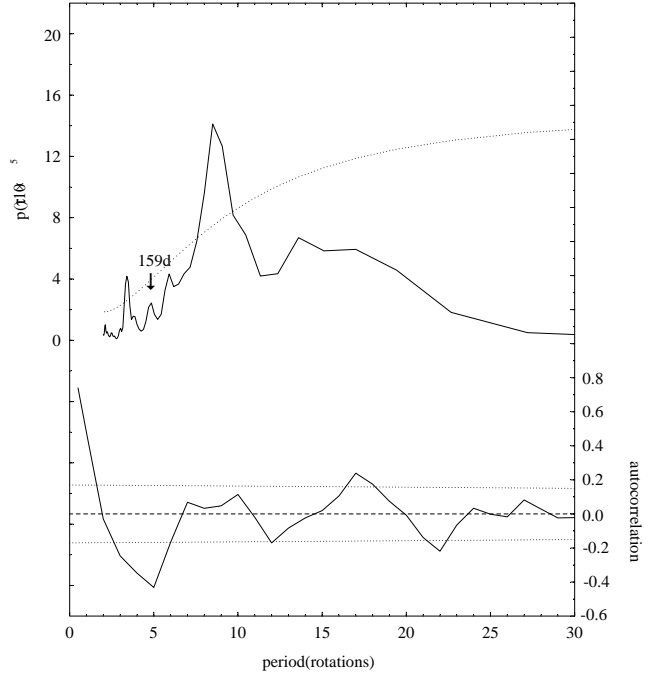}}
\caption{\rm Two upper curves show the periodogram of the time series 
$\{S_n(t_i)-\overline{S_n(t_i)}\}$ (solid line) and the 95\% confidence level 
basing on the'red noise' (dotted line). The periodogram values are presented on 
the left axis. The lower curve illustrates the autocorrelation function of the same 
time series (solid line). The dotted lines represent two standard errors of the 
autocorrelation function. The dashed horizontal line shows the zero level. 
The autocorrelation values are shown in the right axis.}   
   \label{f1}
   \end{figure}

  \begin{figure}
 \centerline{\includegraphics[width=20pc]{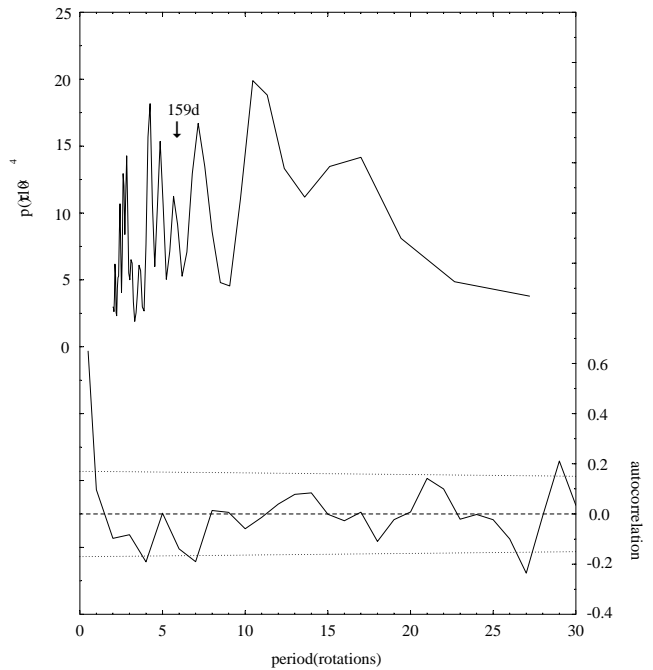}}
  \caption{\rm Same as in Fig.~1, but for the time series 
$\{S_s(t_i)-\overline{S_s(t_i)}\}.$ Because the statistical tests indicate
that the time series is a white noise the confidence level is not marked.}
   \label{f2}
   \end{figure}

 \begin{figure}
 \centerline{\includegraphics[width=20pc]{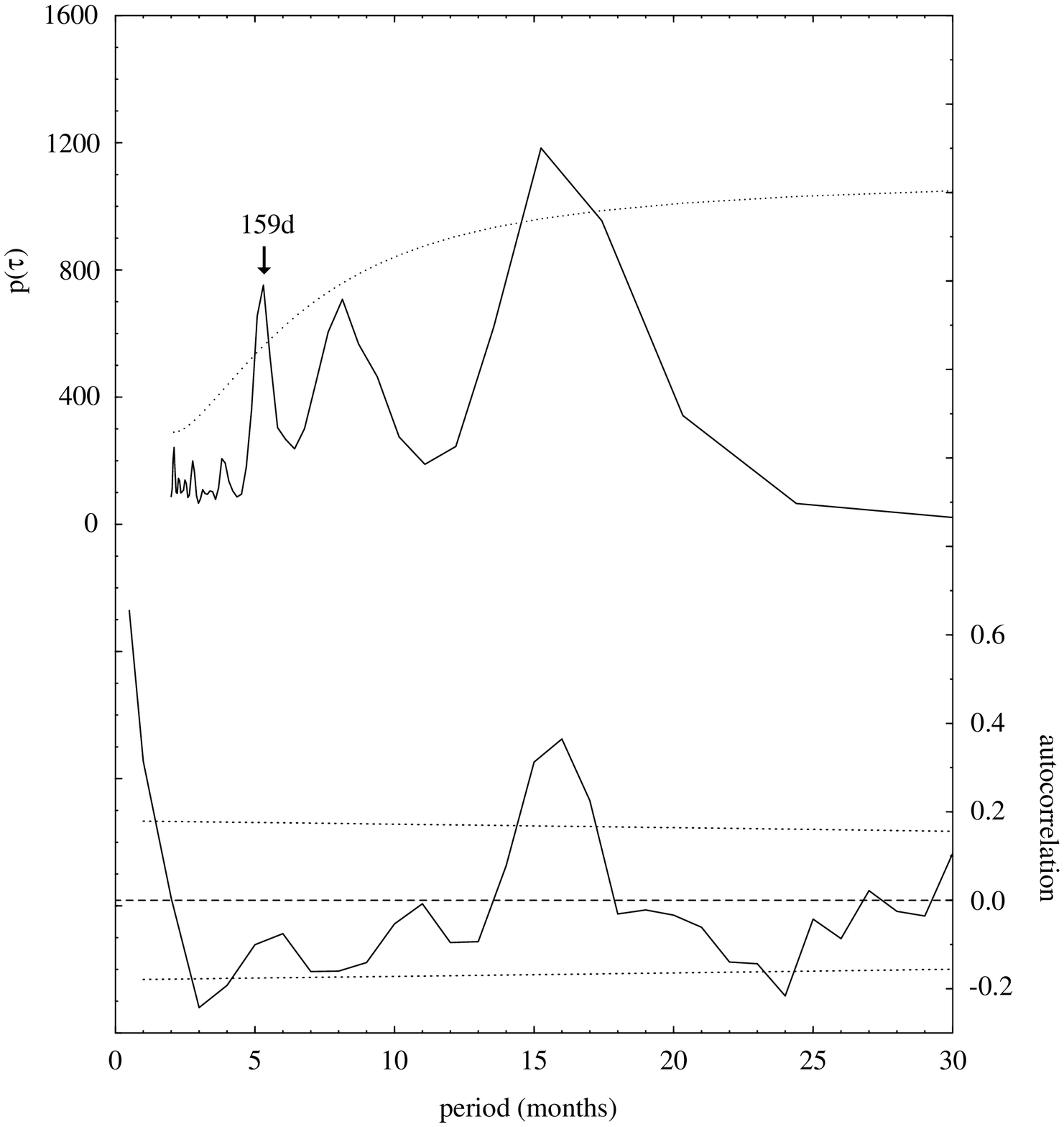}}
  \caption{\rm Same as in Fig.~1, but for the time series $\{R(t_i)-\overline{R(t_i)}\}$.}
   \label{f3}
   \end{figure}

\section{A method of the diagnosis of an echo-effect in the power spectrum}

The method of the diagnosis of an echo-effect in the power spectrum (DE) 
consists in an analysis of a periodogram of a given time 
series computed using the BT method. The BT method bases on 
the cosine transform of the autocorrelation function which creates 
peaks which are in the periodogram, but not in the autocorrelation function.

The DE method is used for peaks which are computed by the FFT method 
(with high resolution) and are statistically significant. 
The time series of sunspot activity indexes with the spacing interval one rotation 
or one month contain a Markov-type persistence, 
which means a tendency for the successive values of the time series to
'remember' their antecendent values. Thus, I use a confidence level 
basing on the 'red noise' of Markov \citep{mit} for the choice of the 
significant peaks of the periodogram computed by the FFT method. 
When a time series does not contain the Markov-type persistence 
I apply the Fisher test and the Kolmogorov-Smirnov test at the significance level 
$\alpha =0.05$ \citep{broc} to verify a statistically significance of periodograms peaks. 
The Fisher test checks the null hypothesis that the time series is white noise agains 
the alternative hypothesis that the time series contains an added deterministic periodic 
component of unspecified frequency. Because the Fisher test tends to be severe in rejecting 
peaks as insignificant the Kolmogorov-Smirnov test is also used.

The DE method analyses 'raw' estimators of the power spectrum. 
They are given as follows

\begin{equation}
\hat{S_k}=\frac{c_0}{M}+\frac{(-1)^k c_M}{M}+\frac{2}{M}\sum_{\tau=1}^{M-1} c_\tau\;cos(\frac{\pi k \tau}{M})
\end{equation}
for $k=1,\ldots ,M,$\\
where
$$c_\tau=\frac{\frac{1}{N-\tau}\sum_{i=1}^{N-\tau}{(X_i-\overline{X})
(X_{i+\tau}-\overline{X})}}{\frac{1}{N}\sum_{i=1}^{N}{(X_i-\overline{X})}^2}$$
for $\tau=1,\ldots ,M,$\\
$N$ is the length of the time series $\{X_i\}$ and $\overline{X}$ is the mean value.

The first term of the estimator $\hat{S_k}$ is constant.
The second term takes two values (depending on odd or even $k$) which
are not significant because $c_M\approx 0$ for large M. Thus, the third
term of (1) should be analysed. Looking for intervals of $\tau$ for which
$\sum_{\tau} c_\tau\;cos(\frac{\pi k \tau}{M})$ has the same sign and different signs
one can find such parts of the function $c_\tau$ which create the value 
$\hat{S_k}$.

Let the set of values of the independent variable of the autocorrelation
function be called $I=\{\tau:\tau=1,\ldots ,M\}$ and it can be divided into the sums 
of disjoint sets:
\begin{equation}
I=\bigcup A_i\;\cup\;\bigcup B_j\;\cup\;\bigcup C_l,
\end{equation}
where\\
$A_0=\{\tau:\;c_\tau\cos(\frac{\pi k \tau}{M})<0 \wedge  0<\tau<\frac{M}{2k}\},$\\
$$A_i=\{\tau:c_\tau\cos(\frac{\pi k\tau}{M})<0 \wedge \frac{(2i-1)M}{2k}<\tau<$$
$$\frac{(2i+1)M}{2k}\},\quad \mbox{for} \quad i=1,\ldots ,k-1,$$
$A_k=\{\tau:\;c_\tau\cos(\frac{\pi k \tau}{M})<0 \wedge (2k-1)\frac{M}{2k}<\tau<M\},$\\[1.5em]
$B_0=\{\tau:\;c_\tau<0 \wedge \cos(\frac{\pi k\tau}{M})<0 \wedge  0<\tau<\frac{M}{2k}\},$\\
$$B_j=\{\tau:\;c_\tau<0\;\wedge\;\cos(\frac{\pi k \tau}{M})<0\;\wedge\;(2j-1)\frac{M}{2k}<$$
\begin{equation}
\quad\tau<(2j+1)\frac{M}{2k}\},\quad \mbox{for} \quad i=1,\ldots ,k-1,\end{equation}
$B_k=\{\tau:\;c_\tau<0 \wedge \cos(\frac{\pi k\tau}{M})<0 \wedge (2k-1)\frac{M}{2k}<\tau<M\},$
\begin{equation}
C_0=\{\tau:c_\tau>0 \wedge \cos(\frac{\pi k\tau}{M})>0 \wedge 0<\tau<\frac{M}{2k}\},
\end{equation}
$$C_l=\{\tau:\;c_\tau>0\;\wedge\;\cos(\frac{\pi k \tau}{M})>0\;\wedge\;(2l-1)\frac{M}{2k}<$$
\begin{equation}
\quad\tau<(2l+1)\frac{M}{2k}\,\},\quad \mbox{for} \quad  l=1,\ldots,k-1,
\end{equation}
$$C_k=\{\tau:\;c_\tau>0 \wedge \cos(\frac{\pi k\tau}{M})>0 \wedge (2k-1)\frac{M}{2k}<$$ 
\begin{equation}
\quad \tau<M\}.
\end{equation}

Well, the set $C_l$ contains all integer values of $\tau$ from the interval of 
$((2l-1)\frac{M}{2k}, (2l+1)\frac{M}{2k})$ for which the autocorrelation function 
and the cosinus function with the period $\Big [\frac{2M}{k}\Big]$ are positive.
The index $l$ indicates successive parts of the cosinus function for which the 
cosinuses of successive values of $\tau$ have the same sign. However,
sometimes the set $C_l$ can be empty. For example,  for $k=13$ and $M=1000$ 
the set $C_1$ should contain all $\tau\in [38, 115]$ for which 
$c_\tau>0$ and $cos(\frac{\pi k \tau}{M})>0$, but for such values of $\tau$ 
the values of $cos(\frac{\pi k \tau}{M})$ are negative. Thus, the set $C_1$ is empty.

\begin{figure}
 \centerline{\includegraphics[width=20pc]{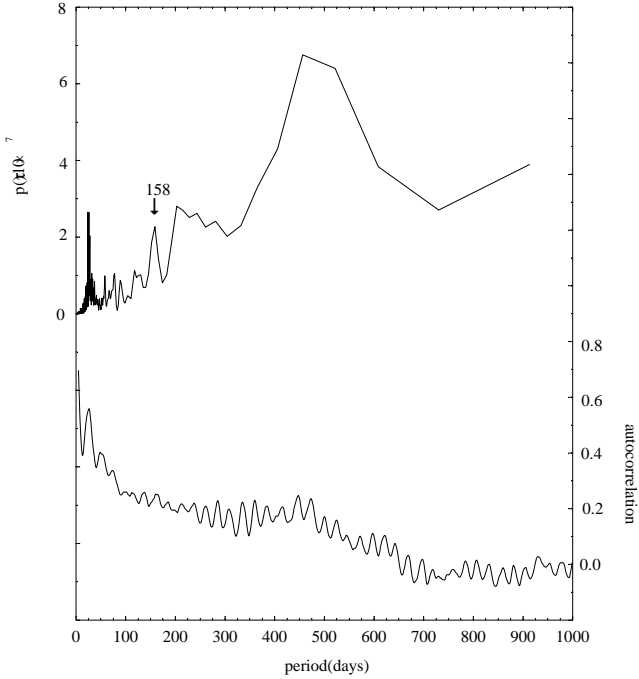}}
  \caption{\rm The upper curve shows the periodogram of the time series 
$\{S^d(t_i)\}$. The periodogram values are presented on 
the left axis. The lower curve illustrates the autocorrelation function of the same 
time series. The autocorrelation values are shown in the right axis.}
   \label{f4}
   \end{figure}

Let us take into consideration all sets \{$A_i$\}, \{$B_j$\} and \{$C_l$\} 
which are not empty. Because numberings and power of these sets depend on the form 
of the autocorrelation function of the given time series, it is impossible to establish 
them arbitrary. Thus, the sets of appropriate indexes of the sets \{$A_i$\}, \{$B_j$\} 
and \{$C_l$\} are called $I_A$, $I_B$ and $I_C$ respectively. 
For example the set $I_C$ contains all $l$ from the set 
$\{1,\ldots ,k\}$ for which the sets $C_l$ are not empty.

To separate quantitatively in the estimator $\hat{S_k}$ the positive contributions which 
are originated by the cases described by the formula (5) from the cases which are described 
by the formula (3) the following indexes are introduced:
$$ds_k^+(l)=\frac{100\;ws_k^+(l)}{ws_k} \quad \mbox{for each } \quad l\in I_C,$$
$$ds_k^-(j)=\frac{100\;ws_k^-(j)}{ws_k} \quad \mbox{for each } \quad j\in I_B,$$
$$ds_k^+=\sum_{l\in I_C} ds_k^+(l),$$
$$ds_k^-=\sum_{j\in I_B} ds_k^-(j),$$
where
$$ws_k^+(l)=\sum_{\tau\in C_l} c_\tau\;cos(\frac{\pi k \tau}{M}) \quad \mbox{for each} \quad l\in I_C,$$
$$ws_k^-(j)=\sum_{\tau\in B_j} c_\tau\;cos(\frac{\pi k \tau}{M}) \quad \mbox{for each} \quad j\in I_B,$$
$$ws_k=\sum_{j\in I_B} ws_k^-(j) +\sum_{l\in I_C} ws_k^+(l),$$
taking for the empty sets \{$B_j$\} and \{$C_l$\} the indices $ds_k^+(l)$ and $ds_k^-(l)$ equal 
zero. 

The index $ds_k^+(l)$ describes a percentage of the contribution of the case 
when $c_\tau$ and $cos(\frac{\pi k \tau}{M})$ are positive to the positive part of 
the third term of the sum (1). The index $ds_k^-(l)$ describes a similar contribution, but for the case 
when  the both $c_\tau$ and $cos(\frac{\pi k \tau}{M})$ are simultaneously negative. 
Thanks to these one can decide which the positive  or the negative values of the autocorrelation 
function have a larger contribution to the positive values of the estimator $\hat{S_k}$.
When the difference $ws_k^+(l)-ws_k^-(j)$
is positive, the statement 'the $k$-th peak really exists' can not be rejected. 
Thus, the following formula should be satisfied:
\begin{equation}
2ds_k^+-100>0.
\end{equation}

Because the $k$-th peak could exist as a result of the echo-effect,
it is necessary to verify the second condition:

\begin{equation}
\exists m\in I_C,\quad ds_k^+(m)=\max_{l\in I_C}
ds_k^+(l) \Rightarrow \Big [\frac{2M}{k}\Big]\in C_m.
\end{equation}

\begin{figure}
 \centerline{\includegraphics[width=20pc]{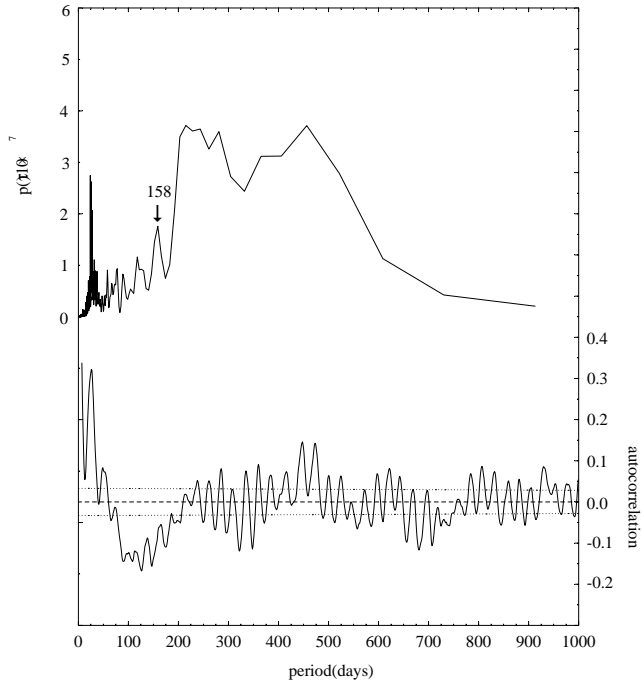}}
  \caption{\rm The upper curve shows the periodogram of the time series 
$\{S^d(t_i)-\overline{S^d(t_i)}\}$. The periodogram values are presented on 
the left axis. The lower curve illustrates the autocorrelation function of the same 
time series (solid line). The dotted lines represent two standard errors of the 
autocorrelation function. The dashed horizontal line shows the zero level. 
The autocorrelation values are shown in the right axis.}
   \label{f5}
   \end{figure}
To verify the implication (8) firstly it is necessary to evaluate the  
sets $C_l$ for $l\in I_C$ of the values of $\tau$ for which the 
autocorrelation function and the cosine function with the period 
$\Big [\frac{2M}{k}\Big ]$ are positive and the sets $B_j, j\in I_B$ 
of values of $\tau$ for which the autocorrelation function and the cosine 
function with the period $\Big [\frac{2M}{k}\Big]$ are negative. 
Secondly, a percentage of the contribution of the sum of products of 
positive values of $c_\tau$ and $cos(\frac{\pi k \tau}{M})$ to the sum 
of positive products of the values of $c_\tau$ and $cos(\frac{\pi k \tau}{M})$
should be evaluated. As a result the indexes $ds_k^+(l)$ for each set $C_l$ 
where $l$ is the index from the set $I_C$ are obtained. Thirdly, from all sets $C_l$
such that $l\in I_C$ the set $C_m$ for which the index $ds_k^+(l)$
is the greatest should be chosen.

The implication (8) is true when the set $C_m$ includes the considered period
$\Big [\frac{2M}{k}\Big]$. This means that the greatest contribution of 
positive values of the autocorrelation function and positive cosines with 
the period $\Big [\frac{2M}{k}\Big]$ to the periodogram value $\hat{S_k}$ 
is caused by the sum of positive products of $(X_i-\overline{X})(X_{i+\tau}-\overline{X})$
 for each $\tau\in([\frac{2M}{k}]-\frac{M}{2k},[\frac{2M}{k}]+\frac{M}{2k})$.

When the implication (8) is false, the peak  $\hat{S_k}$ is mainly 
created by the sum of positive products of $(X_i-\overline{X})(X_{i+\tau}-\overline{X})$ for each 
$\tau\in \Big (\Big [\frac{2M}{n}\Big ]-\frac{M}{2k},\Big [\frac{2M}{n}\Big ]+\frac{M}{2k} \Big )$,
where $n$ is a multiple or a divisor of $k$.

It is necessary to add, that the DE method should be applied to the 
periodograms peaks, which probably exist because of the echo-effect. 
It enables one to find such parts of the autocorrelation function, 
which have the significant contribution to the considered peak.  
The fact, that the conditions (7) and (8) are satisfied, can
unambiguously decide about the existence of the considered periodicity 
in the given time series, but if at least one of them is not satisfied, 
one can doubt about the existence of the considered periodicity. Thus, 
in such cases the sentence 'the peak can not be treated as true' should 
be used.

Using the DE method it is necessary to remember about the power of 
the set $I$. If $M$ is too large, errors of an 
autocorrelation function estimation appear. They are 
caused by the finite length of the given time series and as 
a result additional peaks of the periodogram occur. If $M$ is 
too small, there are less peaks because of a low resolution 
of the periodogram. In applications $M<\frac{N}{3}$ 
is used. In order to evaluate the value $M$ the FFT method is used. 
The periodograms computed by the BT and the FFT method are compared. 
The conformity of them enables one to obtain the value $M$.

 \begin{figure}
 \centerline{\includegraphics[width=20pc]{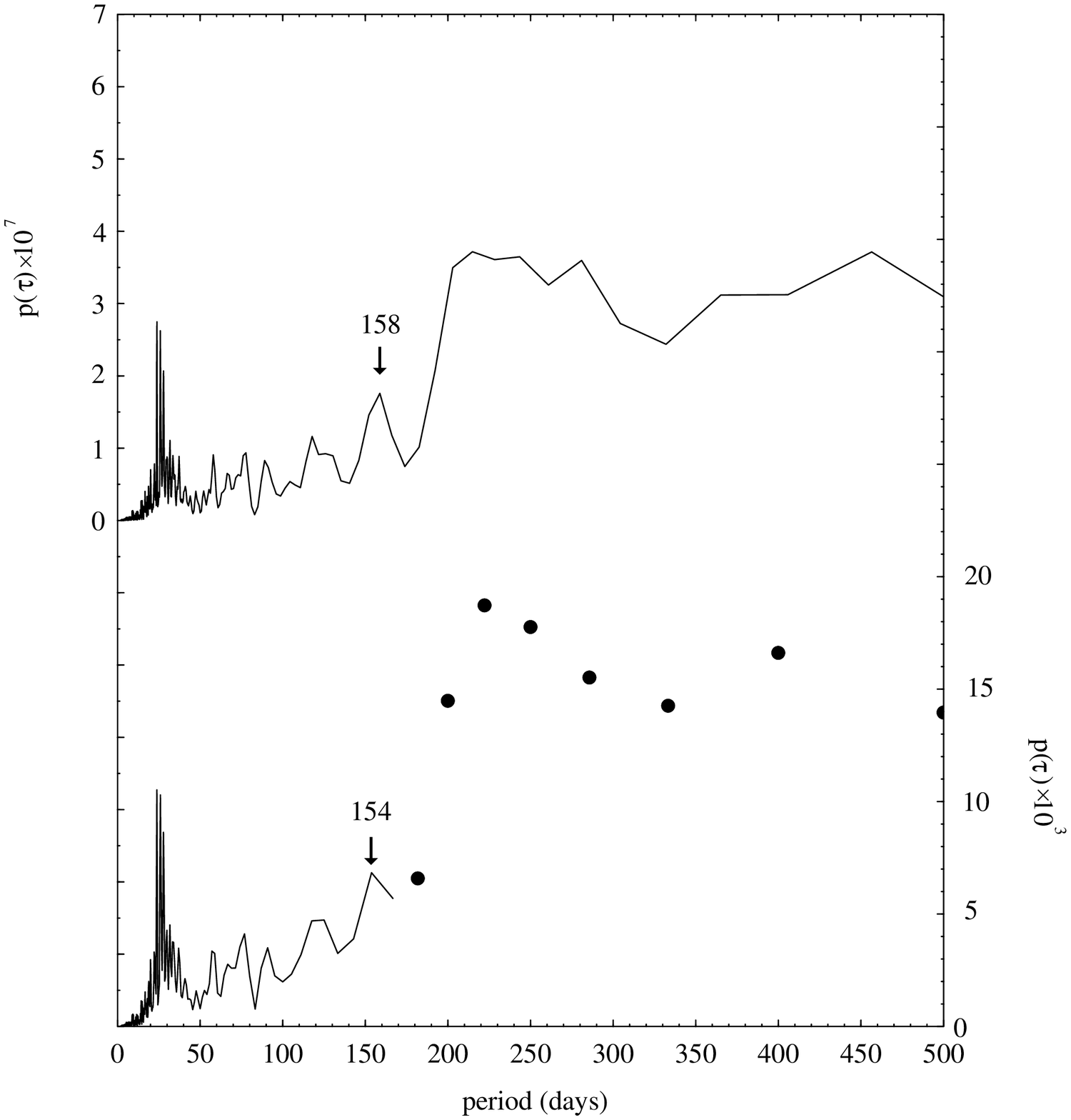}}
  \caption{\rm The upper curve shows the FFT periodogram of the time series 
$\{S^d(t_i)-\overline{S^d(t_i)}\}$. The FFT periodogram values are presented on the left axis. 
The lower curve illustrates the BT periodogram of the same time series 
(solid line and large black circles). The BT periodogram values are shown in the right axis.} 
   \label{f6}
   \end{figure}

\section{Data analysis}

In this paper the sunspot activity data (August 1923 - October 1933)
provided by the Greenwich Photoheliographic Results (GPR)
are analysed. Firstly, I consider the monthly sunspot number data. 
To eliminate the 11-year trend from these data, 
the consecutively smoothed monthly sunspot number $(\overline{R(t_i)})$
is subtracted from the monthly sunspot number $(R(t_i))$ where
the consecutive mean $\overline{R(t_i)}$ is given by
$$\overline{R(t_i)}=\frac{1}{13}\sum_{j=i-6}^{i+6} R(t_j) \quad \mbox{for} \quad i=1,\ldots, N.$$
The values $\overline{R(t_i)}$ for $i=1,\ldots, 6$ and $i=N-6,\ldots, N$ are calculated using 
additional data from last six months of cycle 15 and first six months of cycle 17.

  Because of the north-south asymmetry of various solar indices \citep{viz},
the sunspot activity is considered for each solar hemisphere separately.
Analogously to the monthly sunspot numbers, the time series of sunspot areas in the 
northern and southern hemispheres with the spacing interval $\Delta t=1$ rotation are denoted. 
In order to find periodicities, the following time series are used:\\
$\{S_n(t_i)-\overline{S_n(t_i)}\}$~--~\parbox [t]{54mm}{sunspot area 
fluctuations of the one rotation time interval in the northern hemisphere 
($N=136$),}\\
$\{S_s(t_i)-\overline{S_s(t_i)}\}$~--~\parbox [t]{54mm}{sunspot area
fluctuations of the one rotation time interval in the southern hemisphere 
($N=136$),}\\
$\{R(t_i)-\overline{R(t_i)}\}$~--~\parbox [t]{54mm}{monthly sunspot number fluctuations ($N=122$).}\\ 

In the lower part of Figure \ref{f1} the autocorrelation function of the time series 
for the northern hemisphere $\{S_n(t_i)-\overline{S_n(t_i)}\}$ is shown. It is easy to 
notice that the prominent peak falls at 17 rotations interval (459 days) and $c_\tau$ 
for $\tau\in [3, 6]$ rotations ([81, 162] days) are significantly negative. 
The periodogram of the time series $\{S_n(t_i)-\overline{S_n(t_i)}\}$ 
(see the upper curve in Figures \ref{f1}) does not show the significant peaks at 
$\tau=5, 6$ rotations (135, 162 days), but there is the significant peak at $\tau=9$ 
(243 days). The peaks at $\tau=9, 17$ are close to the peaks of the autocorrelation 
function. Thus, the result obtained for the periodicity at about $155$ days are 
contradict to the results obtained for the time series of daily sunspot areas 
\citep{car2}.

For the southern hemisphere (the lower curve in Figure \ref{f2}) $c_\tau$ for 
$\tau\in [2, 7]$ rotations ([54, 189] days) is not positive except 
$\tau =5$ (135 days) for which $c_5=0.02$ is not statistically significant. 
The upper curve in Figures \ref{f2} presents the periodogram of the time series 
$\{S_s(t_i)-\overline{S_s(t_i)}\}$.  This time series  does not contain 
a Markov-type persistence. Moreover, the Kolmogorov-Smirnov test and the Fisher test 
do not reject a null hypothesis that the time series is a white noise only. This means 
that the time series do not contain an added deterministic periodic component of 
unspecified frequency.

The autocorrelation function of the time series $\{R(t_i)-\overline{R(t_i)}\}$ 
(the lower curve in Figure \ref{f3}) has only one statistically significant peak 
for $\tau = 16$ months (480 days) and negative values for $\tau\in [3,13]$ months 
([90, 390] days). However, the periodogram of this time series (the upper curve in 
Figure \ref{f3}) has two significant peaks the first at 15.2 and the second at 5.3 
months (456, 159 days). Thus, the periodogram contains the significant peak, 
although the autocorrelation function has the negative value at $\tau = 5$ months.

To explain these problems two following time series of daily sunspot areas are considered:\\
$\{S^d(t_i)\}$~--~\parbox [t]{66mm}{daily sunspot areas in the whole solar disk ($N=3653$),}\\
$\{S^d(t_i)-\overline{S^d(t_i)}\}$~--~\parbox [t]{54mm}{daily sunspot area fluctuations,}\\
      where
\begin{equation}
\overline{S^d(t_i)}=\frac{1}{365}\sum_{j=i-182}^{i+182} S^d(t_j) \quad \mbox{for} \quad i=1,\ldots, N.
\end{equation}

The values $\overline{S^d(t_i)}$ for $i=1,\ldots, 182$ and $i=N-182,\ldots, N$ are calculated using 
additional daily data from the solar cycles 15 and 17.

 \begin{figure}
 \centerline{\includegraphics[width=20pc]{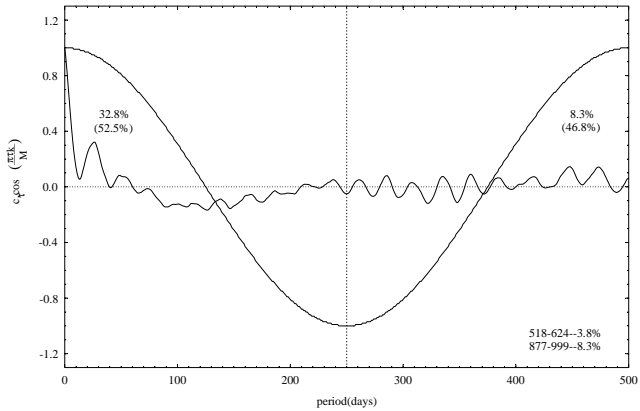}}
  \caption{\rm  Autocorrelation function of the time series $\{S^d(t_i)-\overline{S^d(t_i)}\}$
and the cosine function for $k=13$ (the period at about 154 days). The horizontal line (dotted line) 
shows the zero level. The vertical dotted lines evaluate the intervals where the sets 
$A_i, B_j, C_l$ (for $i, j, l=1,\ldots,k$) are searched. The percentage values show the index 
$ds_k^+(l)$ for each $C_l$ for the time series $\{S^d(t_i)-\overline{S^d(t_i)}\}$ (in parentheses 
for the time series $\{S^d(t_i)\}$). In the right bottom corner the values of $ds_k^+(l)$ for the 
time series $\{S^d(t_i)-\overline{S^d(t_i)}\}$, for $\tau=501,\ldots,1000$  are 
written.}
   \label{f7}
   \end{figure}

 \begin{figure}
 \centerline{\includegraphics[width=20pc]{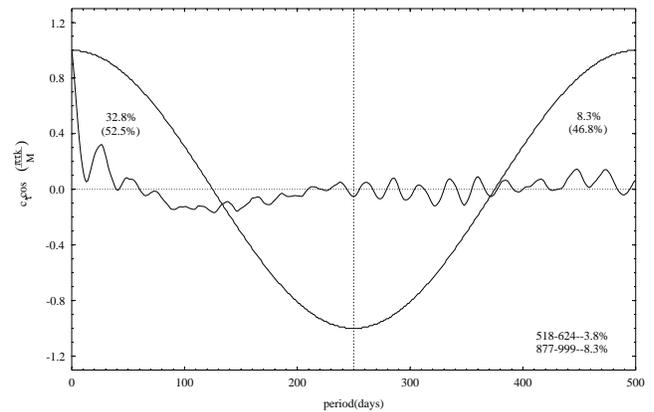}}
  \caption{\rm Same as in Figure 7, but for $k=4$ (the 500-day period )}
   \label{f8}
   \end{figure}

The comparison of the functions $c_\tau$ of the time series $\{S^d(t_i)\}$ 
(the lower curve in Figure \ref{f4}) and $\{S^d(t_i)-\overline{S^d(t_i)}\}$ 
(the lower curve in Figure \ref{f5})
suggests that the positive values of the function $c_{\tau}$ of the time 
series $\{S^d(t_i)\}$ in the interval of $[61,209]$ days could be caused 
by the 11-year cycle. This effect is not visible in the case of 
periodograms of the both time series computed using the FFT method (see the upper 
curves in Figures \ref{f4} and \ref{f5}) or the BT method (see the lower curve in Figure \ref{f6}). 
Moreover, the periodogram of the time series $\{S^d(t_i)-\overline{S^d(t_i)}\}$ has 
the significant values at $\tau=117, 158, 203$ days, but the autocorrelation function is negative 
at these points. \citet{car2} showed that the Lomb-Scargle periodograms 
for the both time series (see \citet{car2}, Figures 7 a-c) have a peak 
at 158.8 days which stands over the FAP level by a significant amount. 
Using the DE method the above discrepancies are obvious. 
To establish  the $M$ value the periodograms computed by the FFT and the BT methods 
are shown in Figure \ref{f6} (the upper and the lower curve respectively). For $M=1000$ 
and for periods less than 166 days there is a good comformity of the both periodograms
(but for periods greater than 166 days the points of the BT periodogram are not linked 
because the BT periodogram has much worse resolution than the FFT periodogram 
(no one know how to do it)). For $M=1000$ and $\tau=154$ the value of $k$ is 13 
($\Big [\frac{2M}{k}\Big ]=153$).
The inequality (7) is satisfied because $2ds_k^+ -100=26\%$. This means that 
the value of $\hat{S_{13}}$ is mainly created by positive values of 
the autocorrelation function. The implication (8) needs an evaluation of 
the greatest value of the index $ds_k^+(l)$ where $l\in I_C$, but the solar 
data contain the most prominent period for $\tau=27$ days because of the solar rotation. 
Thus, although $ds_k^+(0)=\max_{l\in I_C}  ds_k^+(l)$ for each $k<13$, all sets $C_l$
(see (5) and (6)) without the set $C_0$ (see (4)), 
which contains $\tau\in [0,38]$, are considered. 
This situation is presented in Figure \ref{f7}.
In this figure two curves $y=cos\frac{\pi k\tau}{M}$ and $y=c_\tau$
are plotted. The vertical dotted lines evaluate the intervals where the sets 
$A_i, B_j, C_l$ (for $i,~j,~l=1,\ldots,k$) are searched. For such $C_l$ 
two numbers are written: in parentheses the value of $ds_k^+(l)$ for the 
time series $\{S^d(t_i)\}$ and above it the value of $ds_k^+(l)$ for the
time series $\{S^d(t_i)-\overline{S^d(t_i)}\}$. To make this figure 
clear the curves are plotted for the set $\{\tau: \tau=1,\ldots ,500\}$ only. 
(In the right bottom corner information about the values of $ds_k^+(l)$ for the 
time series $\{S^d(t_i)-\overline{S^d(t_i)}\}$, for $\tau=501,\ldots,1000$  are 
written.) 
The implication (8) is not true, because $ds_k^+(m)=14.8\%$ for $m=6$. Therefore, 
$\Big [\frac{2M}{k}\Big]=153\notin C_6=[423,500]$. Moreover, the autocorrelation function 
for $\tau\in [115,192]$ is negative and the set $C_2$ is empty. Thus, $ds_k^+(2)=0\%$. 
On the basis of these information one can state, that the periodogram peak at 
$\tau\approx 154$ days of the time series $\{S^d(t_i)-\overline{S^d(t_i)}\}$ exists 
because of positive $c_\tau$, but  for $\tau$ from the intervals which do not contain 
this period. Looking at the values of $ds_k^+(l)$ of the time 
series $\{S^d(t_i)\}$, one can notice that they 
decrease when $\tau$ increases until $\tau=346$. This indicates, that
when $\tau$ increases, the contribution of the 11-year cycle to the 
peaks of the periodogram decreases. 
An increase of the value of $ds_k^+(l)$ is for $l=6$ for the both time 
series, although the contribution of the 11-year cycle for the time series 
$\{S^d(t_i)\}$ is insignificant. Thus, this part of the autocorrelation 
function ($ds_k^+(6)=\max_{l=1,\ldots ,s} ds_k^+(l)$ for the time series 
$\{S^d(t_i)-\overline{S^d(t_i)}\}$) influences  the $k$-th
peak of the periodogram. This suggests that the periodicity at about 155 days 
is a harmonic of the periodicity from the interval of $[400,500]$ days.
 \begin{figure}
 \centerline{\includegraphics[width=20pc]{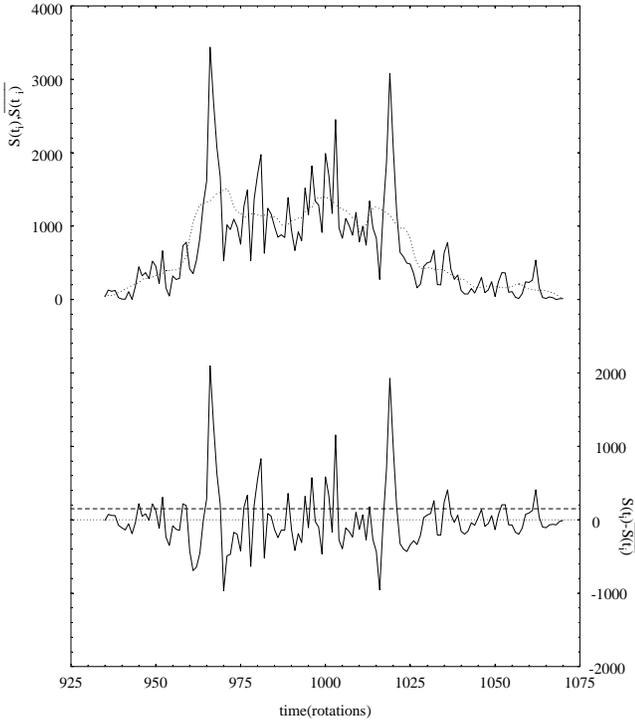}}
  \caption{\rm Two upper curves show sunspot areas of the one rotation time interval 
in the whole solar disk $S(t_i)$ (solid line) and consecutively smoothed  sunspot areas of the 
one rotation time interval $\overline{S(t_i)}$ (dotted line). 
Both indexes are presented on the left axis. The lower curve illustrates
fluctuations of the sunspot areas $S(t_i)-\overline{S(t_i)}$. The dotted and dashed horizontal 
lines represent levels zero and $p$ respectively. The fluctuations 
are shown on the right axis.}
   \label{f9}
   \end{figure}

The described reasoning can be carried out for other values of 
the periodogram. For example, the condition (8) is not satisfied for 
$k=8, 9, 10$ (250, 222, 200 days). Moreover, the autocorrelation function 
at these points is negative. These suggest that there are not a true 
periodicity in the interval of [200, 250] days. It is difficult to decide about 
the existence of the periodicities for $k=6$ (333 days) and $k=7$ (286 days) 
on the basis of above analysis. The implication (8) is not satisfied for $k=7$ and the 
condition (7) is not satisfied for $k=6$, although the function 
$c_\tau$ of the time series $\{S^d(t_i)-\overline{S^d(t_i)}\}$ is significantly positive 
for $\tau=286,333$. The conditions (7) and (8) are satisfied for $k=4$ (Figure \ref{f8}) and $k=5$. 
Therefore, it is possible to exist the periodicity from the interval of $[400,500]$ 
days. Similar results were also obtained  by \citet{lea} for daily sunspot 
numbers and daily sunspot areas. She considered the means of three periodograms 
of these indexes for data from $N=31$ years and found statistically significant peaks 
from the interval of $[400,500]$ (see \citet{lea}, Figure 2).
\citet{kri} studied sunspot areas from 1876-1999 and 
sunspot numbers from 1749-2001 with the help of the wavelet transform. 
They pointed out that the 154-158-day period could be the third harmonic 
of the 1.3-year (475-day) period. Moreover, the both periods fluctuate
considerably with time, being stronger during stronger sunspot cycles.
Therefore, the wavelet analysis suggests a common origin of the both periodicities. 
This conclusion confirms the DE method result which indicates that the periodogram peak at 
$\tau =154$ days is an alias of the periodicity from the interval of $[400,500]$.

\section {The periodicity at about 155 days during the maximum activity period}

In order to verify the existence of the periodicity at about 155 days I consider 
the following time series:\\
$\{S(t_i)-\overline{S(t_i)}\}$~--~\parbox [t]{55mm}{sunspot area 
fluctuations of the one rotation time interval in the whole solar disk 
($N=136$),}\\
$\{S^d_{n}(t_i)-\overline{S^d_{n}(t_i)}\}$~--~\parbox [t]{52mm}{daily sunspot area 
fluctuations in the northern hemisphere from the maximum activity period 
(January 1925 - December 1930, $N=1478$),}\\
$\{S^d_{s}(t_i)-\overline{S^d_{s}(t_i)}\}$~--~\parbox [t]{54mm}{daily sunspot area 
fluctuations in the southern hemisphere from the maximum activity period 
($N=1478$).}\\

\begin{figure}
 \centerline{\includegraphics[width=20pc]{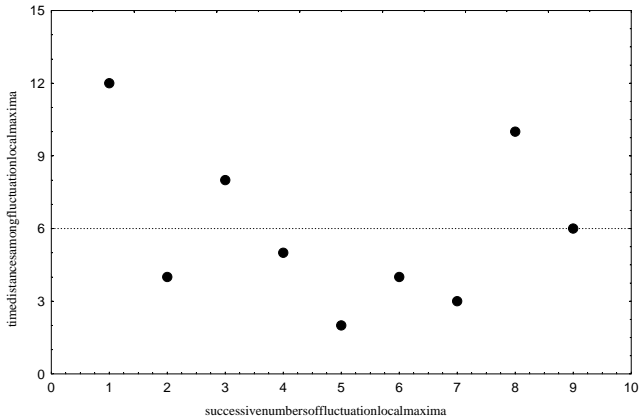}}
  \caption{\rm Nine time distances among the successive local maxima of the values of $F_i>p$. 
  The horizontal line represents the 6-rotation (162-day) period.}
   \label{f10}
   \end{figure}

The value $\overline{S(t_i)}$ is calculated analogously to $\overline{R(t_i)}$
(see Sect. 4). The values $\overline{S^d_{n}(t_i)}$ and $\overline{S^d_{s}(t_i)}$
are evaluated from the formula (9). In the upper part of Figure \ref{f9} the time series 
of sunspot areas $(S(t_i))$ of the one rotation time interval from the whole solar disk 
and the time series of consecutively smoothed sunspot areas 
$(\overline{S(t_i)}=\frac{1}{13}\sum_{j=i-6}^{i+6}{S(t_i)})$ are 
showed. In the lower part  of Figure \ref{f9} the time series of sunspot area fluctuations 
$\{S(t_i)-\overline{S(t_i)}\}$ is presented. On the basis of these data 
the maximum activity period of cycle 16 is evaluated. It is an interval 
between two strongest fluctuations e.a. $J=[966, 1019]$ rotations. 
The length of the time interval $J$ is 54 rotations. If the about 
$155$-day (6 solar rotations) periodicity  existed in this 
time interval and it was characteristic for strong fluctuations from 
this time interval, 10 local maxima in the set of 
$F_i=S(t_i)-\overline{S(t_i)}$ would be seen. Then it should be  
necessary to find such a value of p for which $F_i>p$ for $i\in J$ and the 
number of the local maxima of these values is 10. As it can be seen in the lower part of 
Figure \ref{f9} this is for the case of $p=150$ (in this figure the dashed horizontal 
line is the level of $F_i=150$). Figure \ref{f10} presents 
nine time distances among the successive fluctuation local maxima and 
the horizontal line represents the 6-rotation periodicity. 
It is immediately apparent that the dispersion 
of these points is 10 and it is difficult to find even few points which 
oscillate around the value of 6. Such an analysis was carried out for 
smaller and larger $p$ and the results were similar. Therefore, the 
fact, that the about $155$-day periodicity exists in the time series of 
sunspot area fluctuations during the maximum activity period is 
questionable.

 \begin{figure}
 \centerline{\includegraphics[width=20pc]{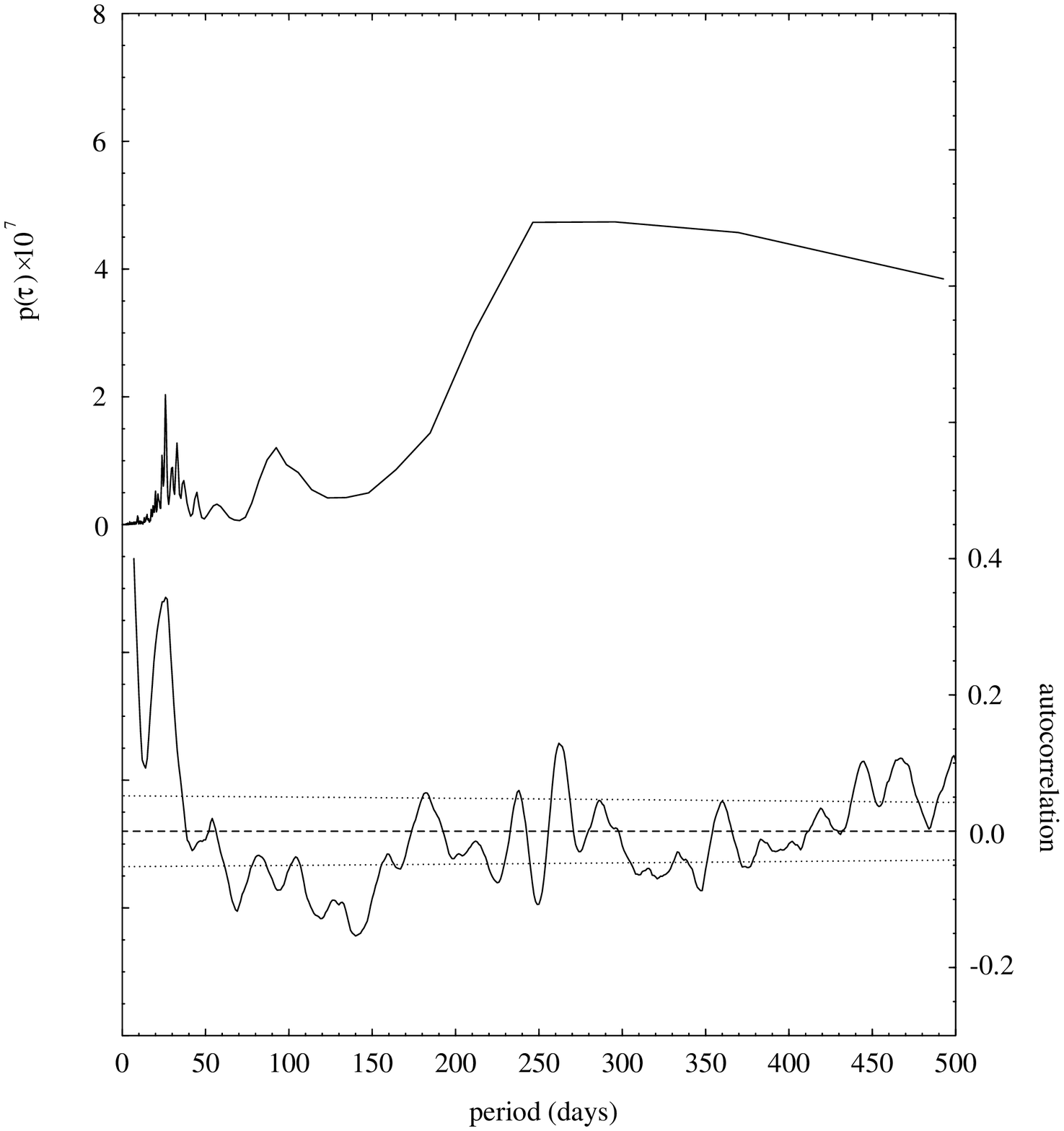}}
  \caption{\rm Same as in Fig.~5, but for the time series $\{S^d_{n}(t_i)-\overline{S^d_{n}(t_i)}\}.$}
   \label{f11}
   \end{figure}

 \begin{figure}
 \centerline{\includegraphics[width=20pc]{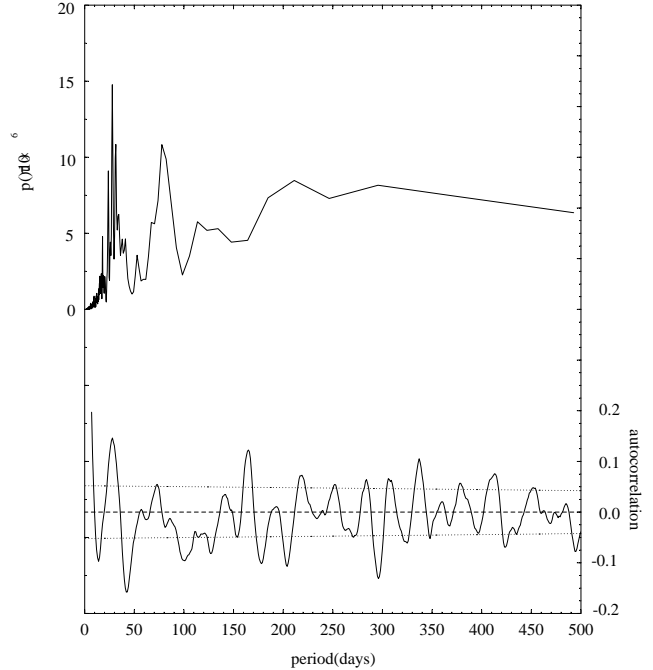}}
  \caption{\rm Same as in Fig.~5, but for the time series $\{S^d_{s}(t_i)-\overline{S^d_{s}(t_i)}\}.$}
   \label{f12}
   \end{figure}

To verify again the existence of the about $155$-day periodicity during the 
maximum activity period in each solar hemisphere separately, the 
time series $\{S_n(t_i)-\overline{S_n(t_i)}\}$ and 
$\{S_s(t_i)-\overline{S_s(t_i)}\}$ were also cut down to the maximum activity 
period (January 1925--December 1930). The comparison of the autocorrelation functions 
of these time series with the appriopriate autocorrelation functions of the time series 
$\{S_n(t_i)-\overline{S_n(t_i)}\}$ and $\{S_s(t_i)-\overline{S_s(t_i)}\}$, which are 
computed for the whole 11-year cycle (the lower curves of Figures \ref{f1} and \ref{f2}), 
indicates that there are not significant differences between them especially for $\tau$=5 
and 6 rotations (135 and 162 days)). This conclusion is confirmed by the analysis of the time 
series $\{S^d_{n}(t_i)-\overline{S^d_{n}(t_i)}\}$ for the maximum activity period.
The autocorrelation function (the lower curve of Figure \ref{f11}) is negative for the interval 
of [57, 173] days, but the resolution of the periodogram is too low to find the significant 
peak at $\tau=158$ days. The autocorrelation function gives the same result as for daily 
sunspot area fluctuations from the whole solar disk ($\{S^{d}(t_i)-\overline{S^{d}(t_i)}\}$) 
(see also the lower curve of Figures \ref{f5}).
In the case of the time series $\{S_s(t_i)-\overline{S_s(t_i)}\}$ $c_5$
is zero for the fluctuations from the whole solar cycle and it is 
almost zero ($c_\tau=0.02$) for the fluctuations from the maximum 
activity period. The value $c_6$ is negative. Similarly to the case of 
the northern hemisphere the autocorrelation function and the periodogram 
of southern hemisphere daily sunspot area fluctuations from the maximum 
activity period $\{S^d_{s}(t_i)-\overline{S^d_{s}(t_i)}\}$ 
are computed (see Figure \ref{f12}).
The autocorrelation function has the statistically significant positive peak 
in the interval of [155, 165] days, but the periodogram has too low resolution to decide 
about the possible periodicities. The correlative analysis indicates that there are 
positive fluctuations with time distances about $155$ days in the maximum activity period.

The results of the analyses of the time series of sunspot area fluctuations from the 
maximum activity period are contradict with the conclusions 
of \citet{lea}. She uses the 
power spectrum analysis only. The periodogram of daily sunspot fluctuations 
contains peaks, which could be harmonics or subharmonics of the true 
periodicities. They could be treated as real periodicities. This effect 
is not visible for sunspot data of the one rotation time interval, but 
averaging could lose true periodicities. This is observed for data from 
the southern hemisphere. There is the about $155$-day peak in the autocorrelation function 
of daily fluctuations, but the correlation for data of the one rotation interval 
is almost zero or negative at the points $\tau=5$ and 6 rotations. Thus, it is 
reasonable to research both time series together using the correlative and the 
power spectrum analyses.

\section{Conclusion}
\vspace{1cm}

The following results are obtained:
\begin{enumerate}
\renewcommand{\theenumi}{(\arabic{enumi})}
\item  A new method of the detection of statistically significant peaks of the periodograms enables 
one to identify aliases in the pe\-rio\-do\-gram.
\item Two effects cause the existence of the peak of the periodogram of the time series 
of sunspot area fluctuations at about $155$ days: the first is caused by the 27-day periodicity, 
which probably creates the 162-day periodicity (it is a subharmonic 
frequency of the 27-day periodicity) and the second is caused by statistically 
significant positive values of the autocorrelation function from the 
intervals of $[400, 500]$ and $[501, 1000]$ days. 
\item The existence of the periodicity of about $155$ days  of the time series of 
sunspot area fluctuations and sunspot area fluctuations from the northern hemisphere 
during the maximum activity period is questionable.
\item The autocorrelation analysis of the time series of sunspot area fluctuations 
from the southern hemisphere indicates that the periodicity of 
about 155 days exists during the maximum activity period.
\end{enumerate}

\section*{Acknowledgments}
I appreciate valuable comments from Professor J. Jakimiec.

\end{document}